\documentclass{article}
\usepackage[utf8]{inputenc}
\usepackage[margin=1in]{geometry} % to change the page dimensions
\usepackage{graphicx} % support the \includegraphics command and options

\usepackage{amsmath}
\usepackage{sectsty}
\allsectionsfont{\sffamily\mdseries\upshape} % (See the fntguide.pdf for font help)
\usepackage{hyperref}
    \hypersetup{
        colorlinks   = true,
        citecolor    = blue,
        linkcolor    = blue
    }

\usepackage[dvipsnames]{xcolor}

\usepackage{titling}

\usepackage{cite}

\title{Enhanced visibility through microbubble-induced photoacoustic fluctuation imaging}
\small
\author{Marco A. Inzunza-Ibarra$^1$ \and J. Angel Navarro-Becerra$^1$ \and Venkatalakshmi Narumanchi$^2$ \and Nick Bottenus$^{1,3}$ \and Todd W. Murray$^{1,3}$ \and Mark A. Borden$^{1,3}$}
\date{%
    $^1$Department of Mechanical Engineering, University of Colorado Boulder, 80309\\%
    $^2$Department of Electrical Engineering, University of Colorado Boulder, 80309\\
    $^3$Biomedical Engineering Program, University of Colorado Boulder, 80309\\[2ex]%
    \today
}

\begin{document}
\maketitle
\begin{abstract}
	A photoacoustic contrast mechanism is presented based on the photoacoustic fluctuations induced by microbubbles flowing inside a micro vessel filled with a continuous absorber. It is demonstrated that the standard deviation of a homogeneous absorber mixed with microbubbles increases non-linearly as the microbubble concentration and microbubble size is increased. This effect is then utilized to perform photoacoustic fluctuation imaging with increased visibility and contrast of a blood flow phantom.  
\end{abstract}

\normalsize
\section{\label{sec:1} Introduction}
Photoacoustic (PA) imaging is based on the PA effect where ultrasound signals are generated by the transient absorption of light and subsequent thermo-elastic expansion \cite{Beard}. The biomedical applications using photoacoustics have grown significantly in recent years due to intrinsic optical absorption sensitivity at the molecular level \cite{Kim2010, Weber,Gujrati}. Some of these applications have significant and widely translatable implications, which include PA molecular imaging \cite{Wilson,Zackrisson,Liu2016,Yao2018}, cancer detection and diagnostics \cite{Mallidi2011,Herzog,LinSingle2018}, and photodynamic \cite{Dougherty,Hester} and photothermal \cite{Shah,Liu2019} cancer therapy. One of the key advantages of PA imaging is the optical wavelength-dependent absorption of endogenous chromophores found in blood, including oxygenated and deoxygenated hemoglobin, making it ideal for angiography \cite{Wray2019}. Moreover, imaging of the blood vessels with high spatial resolution is extremely important in the detection of angiogenesis sites located in the vicinity of tumors \cite{LinSingle2018}. PA tomography (PAT) systems \cite{Li2009} typically employ an array of ultrasound transducers placed outside the object, and images are formed by applying reconstruction algorithms on the detected signals. 

In many practical applications of PAT, access to the imaging object is severely restricted and signals cannot be collected from large detection angles, giving rise to the “limited-view” problem \cite{Xu_recons,Paltauf,Dean_part}. Most commonly, conventional linear array detectors are employed which lack visibility of structures due to a finite-aperture and coherence artifacts generated from PA signals from sub-acoustic  wavelength absorbers \cite{Guo1}. De\'{a}n-Ben and Razansky \cite{Dean1} provided a useful link between PA images acquired under the limited-view scenario as well as the coherent behavior of PA signals from sub-acoustic absorbers. In brief, finite-aperture PA images from objects containing high absorber densities are devoid of structural information  belonging to the interior of objects due to destructive interference, and only the edges associated with constructive interference are retained. 

In the recent years, PA fluctuation imaging has demonstrated to enhance visibility and spatial resolution using multiple speckle illumination \cite{Gateau,ChaigneSpeck,Hojman,MurraySpeck}, SOFI inspired statistical approaches \cite{Dertinger,ChaignePart}, and localization \cite{Betzig,Hess,VilovLoc,DeanLoc,ZhangLoc}. The last two examples demonstrated that the PA fluctuations elicited from flowing absorbers could be leveraged to provide structural information about the object. Flow-induced PA fluctuations, however, can be difficult to detect in blood, where partial waves emitted from densely packed red blood cells interfere destructively and can be perceived as a homogeneous signal at the ultrasound transducer. Recent work by Vilov et al. \cite{VilovFluc} provided a comprehensive theory on PA fluctuation imaging and experimental results demonstrating enhanced visibility using blood at physiological concentrations. 

Since their initial regulatory approval for echocardiography in the mid 1990’s, microbubbles (MBs) have been used routinely as ultrasound contrast agents for a variety of clinical applications \cite{Frinking3Dec}. Photoacoustics is highly sensitive to small changes in optical absorption variations, meaning that a given percent change in the optical absorption coefficient yields the same percentage change in the PA amplitude \cite{WangTut}. Acting as weak absorbers, randomly distributed MBs in the excitation volume would hypothetically induce fluctuations in the absorption profile and PA response. Here, we experimentally demonstrate increased imaging visibility and contrast under limited-view conditions using PA fluctuation imaging induced by MBs. 

\section{\label{sec:2} Methods}
Consider the PA response from a homogeneous absorber (ink) and MBs flowing through a cylindrical vessel as shown in Fig. \ref{fig:exp1}(a). Laser pulse illumination excites a PA response from the cylinder and the detected signal by the ultrasound transducer is the sum of the signals generated by the absorbers surrounding the MBs.  Assuming that the optical absorption from the MBs is negligible, they therefore do not emit partial PA waves to contribute to the overall signal. Instead, they interfere with the absorption properties of the object therefore developing a heterogenous absorption profile. Since ink is a homogeneous absorber by itself, we attribute the heterogeneity in the overall PA response to the random distribution of transparent MBs. 

\subsection{Single-element transducer experiments}
To measure the PA response from a homogeneous sample and MBs, ink and MBs were flowed through a 200 $\mu$m inner diameter tube using a syringe pump. Fig. \textcolor{blue}{S1} depicts the experimental setup. The flow speed is set to an average velocity of 10 mm/s to ensure that the spatial distributions of MBs in adjacent single-shots are uncorrelated. A  micro-chip diode pumped laser with one nanosecond laser pulses firing at a pulse repetition frequency (PRF) of 1 kHz and 532 nm wavelength is used to generate the PA signals. Using a F = 150 mm lens, we set the laser spot-size to 50 $\mu$m and the laser fluence was set to 20 mJ/cm$^2$. A single-element focused ultrasound transducer (Olympus V317) with a 20 MHz center frequency and a receive-only full-width at half-maximum spot-size of 208 $\mu$m was used to detect the PA waves. A microscope glass cover-slip was used as an acoustic reflector in order to align both the acoustical and optical focal spots. A 1 GHz amplifier (FEMTO, HSA-X) with 40 dB gain was used to amplify the radio-frequency (RF) signals detected by the single-element transducer. The PA signals were acquired using a high-speed digitizer at 2.5 GHz sampling frequency and were post-processed by a low-pass Butterworth filter with a 35 MHz cutoff frequency.

\begin{figure}
    \centering
    \includegraphics[width=\linewidth]{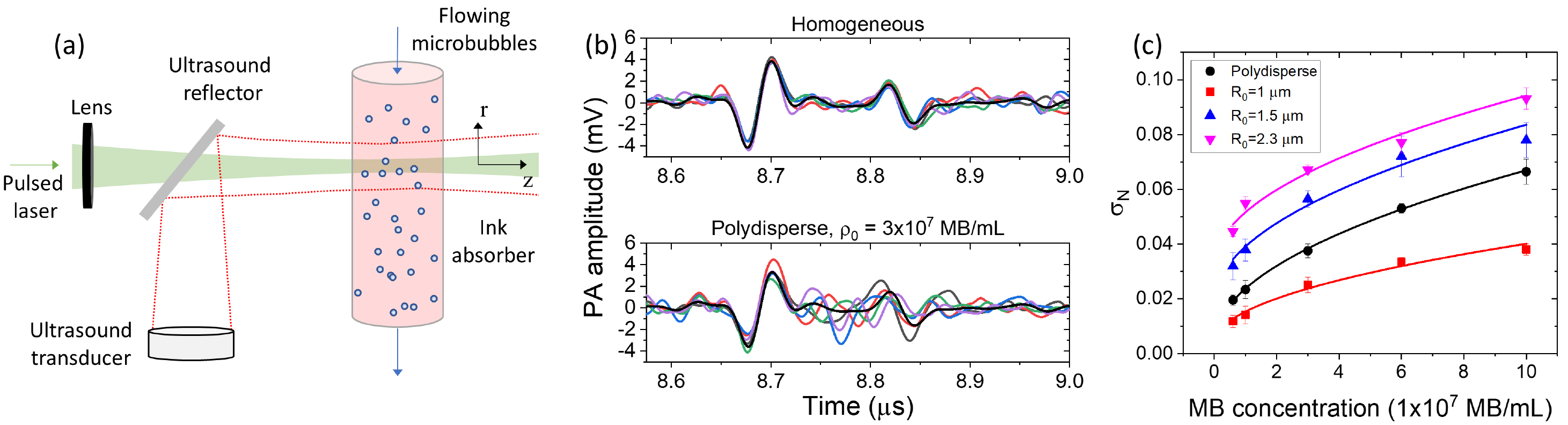}

    \caption{MB induced PA fluctuation measurements. (a) Experimental concept and geometry. (b) In color are five PA single-shots for the homogeneous sample (ink) and the homogeneous sample plus MBs with a polydisperse distribution and concentration $\rho_0$=3×10$^7$ MB/mL. In black are the mean signals of the 5000 single-shots. (c) Normalized standard deviation $\sigma_N$ as a function of MB concentration for four different MB size distributions.}
    \label{fig:exp1}
\end{figure}

Five thousand single-shot PA signals were collected in order to compute the mean and standard deviation over time. Five PA single-shots along with the mean response (black curve) for the homogeneous case (ink), as well as ink and polydisperse MBs of concentration $\rho_0$=3x10$^7$ MB/mL with diameters ranging from 1-10 $\mu$m, are shown in Fig. \ref{fig:exp1}(b). The homogeneous case demonstrates variations in the single-shots that are associated with the electronic and system noise. A notable increase in the single-shot variations above the system noise is seen in the ink sample with the polydisperse MB distribution. The fluctuations cancel in the mean response for both cases as only the edges of the object, the front and back of the micro vessel, are observed. We also note high variations in the single-shots around the back wall of micro vessel. We quantify the fluctuation level in the five thousand PA single-shots by computing the normalized standard deviation $\sigma_N$, expressed as $\sigma_N=\sqrt{\sigma_m^2-\sigma_s^2}/A_{pp}$ \cite{Inzunza}.

The peak-to-peak amplitude $A_{pp}$ is computed from the mean response as a function of time. Similarly, the measured standard deviation (SD) $\sigma_m$, is taken as the average value from a 50 ns time window centered at 8.75 $\mu$s in the SD versus time plot. The system SD $\sigma_s$, is a fixed value of the SD versus time before the PA arrival of the micro vessel. $\sigma_N$ was measured for different MB concentrations and size distributions, as shown in Fig. \ref{fig:exp1}(c). The preparation and characterization of the MBs are described in more detail in the supplemental material and Fig. \textcolor{blue}{S2}. In Fig. \ref{fig:exp1}(c), we demonstrate a nonlinear increase of the $\sigma_N$ as a function of concentration. As ink responds as a homogeneous medium, the increase in $\sigma_N$ can be attributed exclusively to the MBs as they perturb the absorption profile. Interestingly, a square-root regression was applied to $\sigma_N$ versus MB concentration, shown as the straight lines in Fig. \ref{fig:exp1}(c), with all the fits having an adjusted R$^2$ value greater than 0.95. 
 
This square-root dependence of $\sigma_N$ is inversely related to fluctuations induced by absorbers. In previous works \cite{Guo1,Dean1,VilovFluc,Inzunza}, it has been reported that the PA fluctuations arising from the interior of micro-vessels decrease as the absorber concentration increases and follows an inverse square-root dependence. Fundamentally , this makes sense based on previous work in that an increase in the density of the microbubbles leads to a decrease in the density of absorbers. 

A significant increase in $\sigma_N$ was also observed as we increased the average MB radius. For example, the fluctuation level $\sigma_N$ for the concentration of 6x10$^7$ MB/mL is about 230\% higher for R$_0$ = 2.3 $\mu$m compared to R$_0$ = 1 $\mu$m. This is an important result as higher fluctuation levels are preferable as we move towards PAT with acoustic resolution, where the excitation and detection regions are significantly larger than in the experiment described here. This result motivates the next section of our study in which we utilized the fluctuations generated by MBs in ink as well as blood to perform PA fluctuation imaging.

\subsection{Photoacoustic fluctuation imaging experiment}
A PA imaging system was built as shown in Fig. \textcolor{blue}{S3}. A 20 MHz center frequency linear array (128 elements, L22-14vx) with a 60 percent bandwidth at -6 dB and an elevational focus at 8 mm was used to receive the PA RF signals. The probe was connected to an ultrasound scanner (Verasonics Vantage) where the 128-channel raw frames sampled at frequency $f_s$ = 62.5 MHz were sent via PXIe to a host computer. Laser pulses from an optical parametric oscillator (OPO) tuned at 688 nm wavelength, pumped by a Q-switched laser (Surelite I-20, PRF = 20 Hz, 7 ns pulse-width at 532 nm wavelength), were used to illuminate a micro vessel phantom. A knife-edge measurement was made to determine the 1/e optical spot-size of 7.5 mm at the micro vessel location, and the fluence was set to 20 mJ/cm$^2$. The micro vessel phantom was surgical tubing (Scientific Commodities) made from low-density polyethylene with an inner diameter of 0.58 mm. The linear array was aligned so that the direction of the transducer elements was perpendicular to the micro vessel as shown in Fig. \textcolor{blue}{S3}(b). Finally, a syringe pump was used to control the fluid flow speed, which was set to an average velocity of 15 mm/s, providing uncorrelated MB spatial distributions for every laser shot. Additionally, a stir plate was used to mix the absorber and MB solution before flowing into the vessel. In the previous section, we characterized the MB induced PA fluctuations $\sigma_N$ for different MB size distributions. Based on those results, we chose the size-isolated MBs with an average radius of R$_0$ = 2.3 $\mu$m, which resulted in the largest $\sigma_N$. The protocol for MB synthesis and size isolation is described in the supplemental material. The measured size distribution for R$_0$ = 2.3 $\mu$m for the imaging experiments was similar to that in Fig. \textcolor{blue}{S2}. 

The PA RF frame acquisition was triggered by sampling part of the 688 nm beam onto a photodiode. The photodiode then triggered a digital delay generator, which provided the input trigger TTL signal to the ultrasound scanner. In order to compensate for the jitter between the laser pulse and PA acquisition, an output sync signal from the ultrasound scanner was recorded on a second computer along with the laser pulse. The precise difference in time between the laser pulse and the PA acquisition was used to shift the frames in time. Each frame consists of 128 time resolved RF signals, and an ensemble of 1,000 frames was acquired in order to generate the mean and SD images. A conventional delay-and-sum (DAS) beamforming algorithm with an axial and lateral  step-size of 10 $\mu$m was performed on the Hilbert transformed RF data to produce the reconstructed in-phase and quadrature (IQ) PA images. A flow diagram of the image reconstruction steps is provided in supplemental Fig. \textcolor{blue}{S4}. 

Parasitic noise in the IQ frames associated with fluctuations in the laser energy, for example, can arise as dominating features in the fluctuation images. Therefore, spatiotemporal filtering using singular value decomposition (SVD)  was a necessary step in our data processing. A detailed description of the SVD processing is provided in the supplemental material. Briefly, the SVD of the images can be written as $\boldsymbol{S=U\Sigma V^T}$, where the first singular values in $\boldsymbol{\Sigma}$ correspond to highly coherent spatial  features in the IQ images, such as the object’s edges. We therefore used the SVD to suppress the highly correlated features by setting the first singular values in $\boldsymbol{\Sigma}$ equal to zero. We then computed a new set of images $\boldsymbol{S}$ and calculated the SD, resulting in the final PA fluctuation images presented here.

\subsection{Imaging results}
\begin{figure}[htbp]
    \centering
    \includegraphics[width=\linewidth]{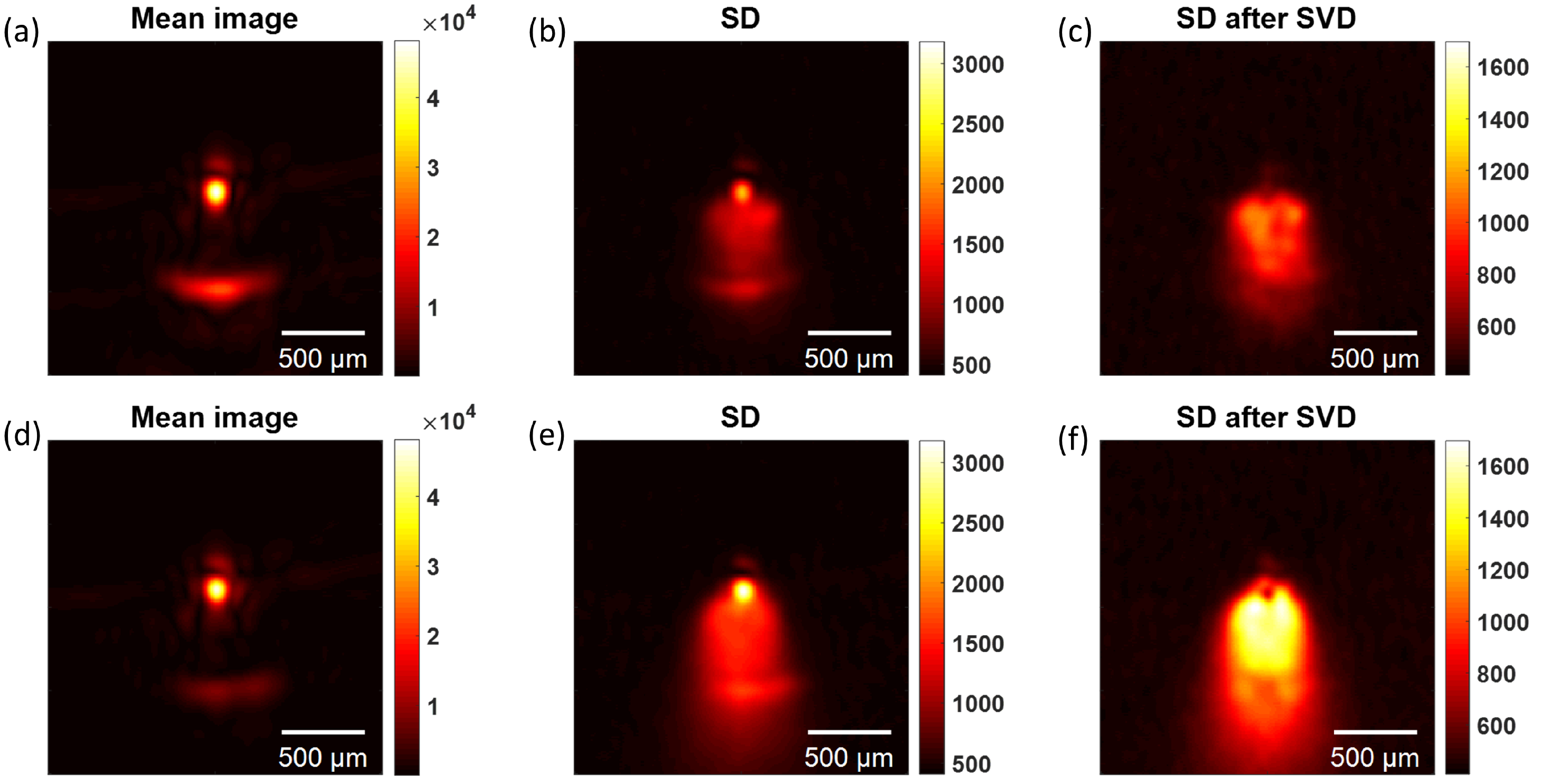}
    \caption{\label{fig:exp2}Reconstructed photoacoustic images for ink and MBs. (a-c) Images of only ink of the mean, SD and SD after SVD, respectively. (d-f) Images for ink and MB concentration of $\rho_0$=6×10$^6$ MB/mL and radius R$_0$ = 2.3 $\mu$m. For both cases, N = 1,000 frames were acquired to compute the statistics and the first two singular values were set to zero in the SVD.}
\end{figure}

The imaging results from the experiment can be seen in Fig. \ref{fig:exp2}. The top row (a-c) shows the resulting images from ink, which served as a homogeneous absorber control sample. The bottom row (d-f) depicts the resulting images from a mixture of ink and size-isolated MBs (similar distribution as shown in Fig. \textcolor{blue}{S2}) of mean radius R$_0$ = 2.3 $\mu$m and concentration $\rho_0$=6×10$^6$ MB/mL. The absolute value of the mean images (Fig. \ref{fig:exp2}(a) and (d)) for both samples were dominated by the limited-view artifact, in which only the front and back edges of the object were resolved, and the interior of the object was hidden. The SD image in both cases (Fig. \ref{fig:exp2}(b) and (e)) demonstrated a notable increase in the visibility through the interior of the object, but the images were mostly dominated by a strong fluctuation feature associated with laser fluctuations at the object’s front wall. This feature was suppressed in the SD after SVD images in Fig. \ref{fig:exp2}(c) and (f). We note that the first two singular values in the SVD were set to zero. We expected that the SD images of the homogeneous sample would not produce an increased noise level, but this was not the case in our results. We attribute this non-zero SD level of the homogeneous sample to variations of the laser pulse energy, which were not entirely compensated for by the SVD. However, as expected from our single-element transducer experimental results, MBs can be leveraged to induce PA fluctuations resulting in increased image contrast well above that of the homogeneous control sample. Additional clutter can be seen after the object in Fig. \ref{fig:exp2}(f) which is associated with single or multiple scattering off the microbubbles of the generated photoacoustic waves.

\begin{figure}
    \centering
    \includegraphics[width=\linewidth]{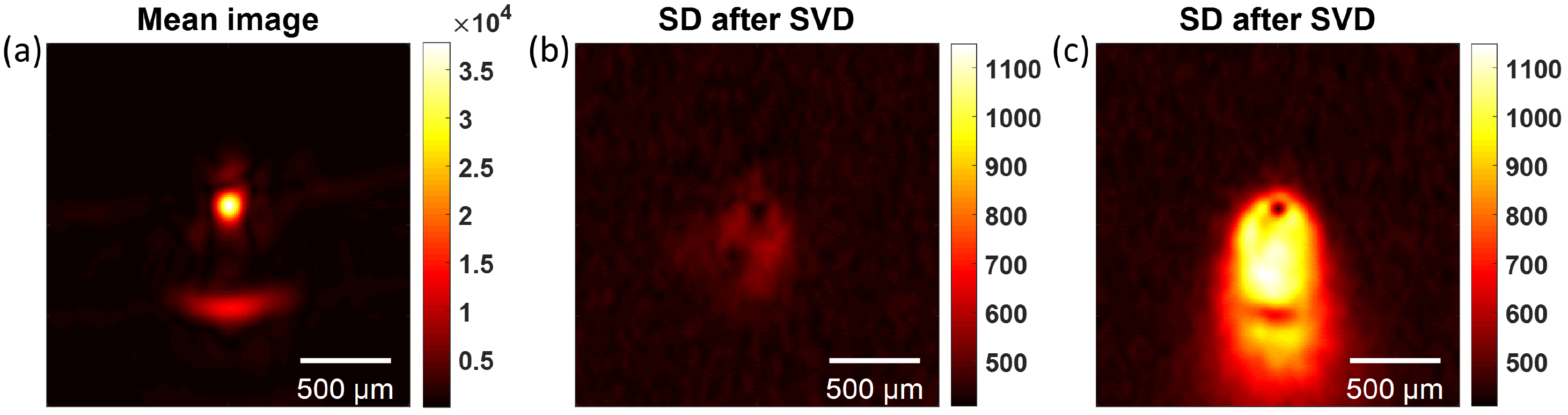}
    \caption{\label{fig:exp3}Reconstructed images for blood and MBs. (a-b) Mean and SD images (after SVD) for blood. (c) SD (after SVD) of blood and MBs $\rho_0$=6×10$^6$ MB/mL and R$_0$ = 2.3 $\mu$m.}
\end{figure}

PA mean and SD images using blood and MBs were also acquired using the same experimental conditions. Undiluted, defibrinated bovine blood (Carolina Biological) with a hematocrit Hct = 0.46 mixed with MBs was prepared at room temperature. For these experiments, we again used size-isolated MBs with a mean radius of R$_0$ = 2.3 $\mu$m. The resulting images for blood and a mixture of blood and MBs at a concentration of $\rho_0$=6×10$^6$ MB/mL are depicted in Fig. \ref{fig:exp3}. Again, N = 1,000 RF frames were acquired to compute the mean and SD (after SVD), and the first two singular values in the SVD were set to zero. The mean and SD (after SVD) images of blood are shown in Fig. \ref{fig:exp3}(a-b), respectively. The SD (after SVD) for the blood and MBs mixture is shown in Fig. \ref{fig:exp3}(c). A notable increase in the visibility of the object’s interior is seen in the SD (after SVD) image for blood and MBs when compared to blood by itself. The mean image of blood is dominated by the limited-view artifacts and the interior of the object is hidden, like the case of blood and MBs (mean not shown here). 

\begin{figure}
    \centering
    \includegraphics[scale=0.5]{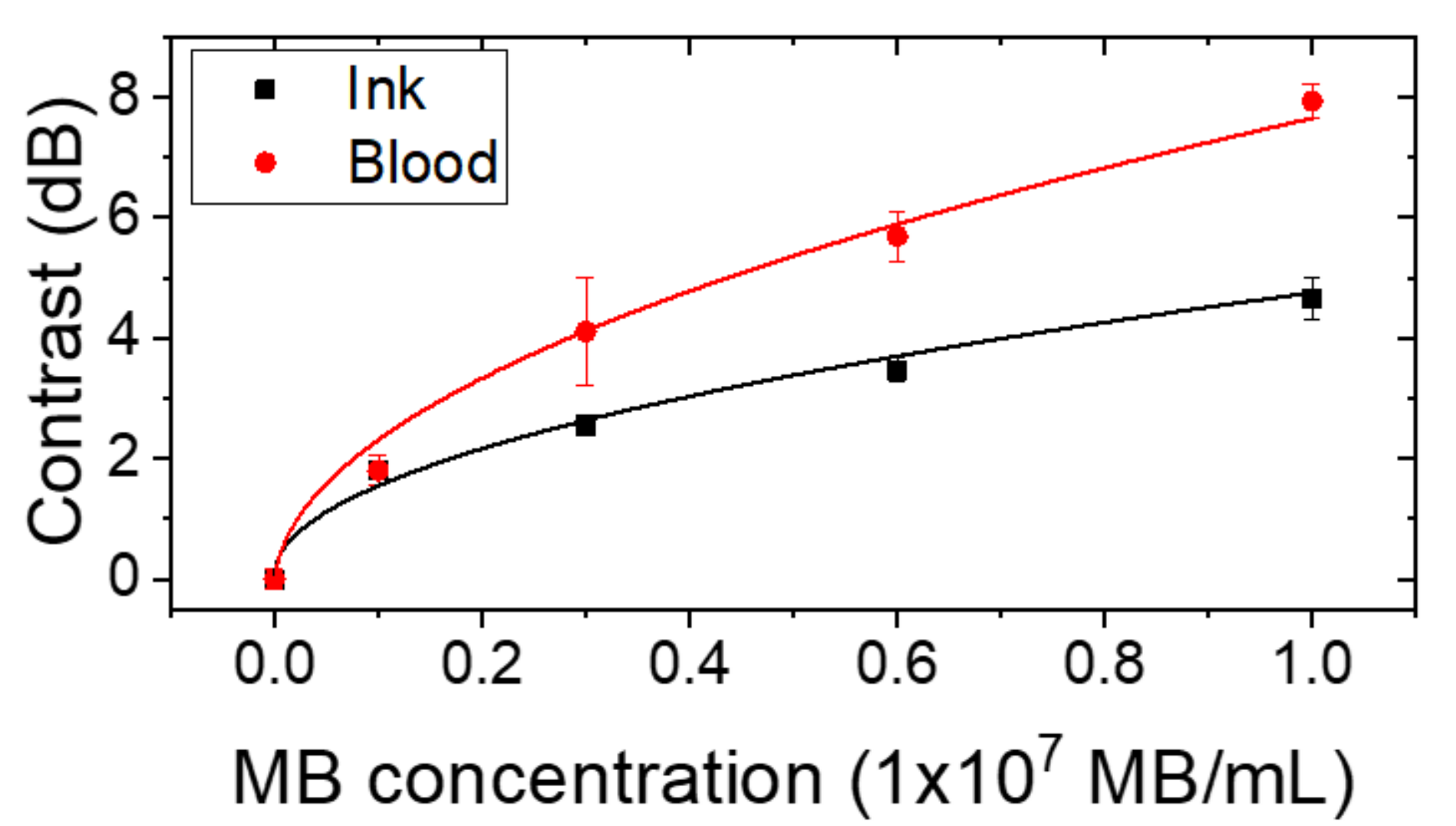}
    \caption{\label{fig:exp4} Contrast as a function of MB concentration for ink and blood samples. Error bars represent the SD of three measurements.}
\end{figure}

In order to quantitatively demonstrate the fluctuation image contrast enhancement as a function of MB concentration, we took images using four different concentrations in ink and blood. The contrast $C$ induced by the MBs was calculated on the SD (after SVD) images using the equation $C=20\log_{10}(\mu_{target}/\mu_{noise})$ where $\mu_{target}$ and $\mu_{noise}$ are the mean pixel intensity inside a region of interest (ROI) of our target object and of the noise background, respectively. The ROI for the target was chosen to be a 0.58 mm diameter circle with an origin centered about the cross-sectional area of our micro-vessel. The mean pixel intensity for the same ROI was calculated for the noise on the top-left corner of our images, where PA contributions from the target were negligible. In Fig. \ref{fig:exp4}, we plot the contrast as a function of MB concentration. The contrast value for the control samples of ink and blood were subtracted from the contrast with MBs. The square markers in black represent the contrast of ink, while the circle markers in red represent contrast of blood. We also plot the SD error bars for three measurements using 1,000 frames per measurement. The contrast for both the ink and blood absorbers increased as a function of MB concentration. Similar to our experimental results in Fig. \ref{fig:exp1}(c), the fluctuation level followed a square-root dependence with MB concentration. The solid lines are a square-root regression fit to the contrast parameter with over 0.95 adjusted R$^2$ values. Remarkably, blood with a MB concentration of $\rho_0$=1×10$^7$ MB/mL expresses a difference in contrast of about 8 dB, or about 250\%, compared to blood alone. 

Introducing MBs in blood to produce incoherence in the PA signals can also be beneficial to improve the accuracy of Doppler based flow velocimetry techniques, for example \cite{Brunker1,Brunker2}, since these techniques rely on high spatiotemporal incoherence for accuracy. However, some of the limitations in combining MBs with photoacoustics is that MBs will inherently exacerbate light scattering through the volume as well as scatter the PA waves propagating towards the detector. As shown in our mean image in Fig. \ref{fig:exp2}(d), the introduction of MBs slightly reduces the amplitude of the coherent feature corresponding to the back of the object. This could possibly lead to deleterious resolution effects when imaging larger objects or objects placed deep in tissue. Moreover, the clutter artifact in our fluctuation images caused by acoustic scattering from the MBs can potentially be compensated for and is the direction of our future work. Nevertheless, we have demonstrated an interesting result which is that an increase in the photoacoustic image contrast can be achieved by MB concentrations as low as 1×10$^6$ MB/mL in blood, which is already a standard dose for contrast-enhanced ultrasound imaging in humans.

\section{\label{sec:3} Conclusion}
In conclusion, we have demonstrated a novel PA image contrast technique in vitro by exploiting the signal fluctuations induced by conventional lipid-coated MBs. The PA fluctuations of different MB concentrations and size distributions were characterized experimentally. We found that the fluctuation level $\sigma_N$ increases with MB concentration as well as size. Finally, fluctuation imaging was performed to overcome the limited-view problem, and it was demonstrated that MBs provided an increase in the contrast and visibility of a blood-filled micro-vessel phantom.

%% If only one acknowledgment, 
%% \begin{acknowledgment}...\end{acknowledgment}
%% may be used.
\section{Acknowledgements}
This research was funded by NIH grant R01CA195051 to MAB. We thank Dr. Francesco Guidi for his advice on the instrumentation and fruitful discussions.

\newpage
\title{Supplemental material for Enhanced visibility through microbubble-induced photoacoustic fluctuation imaging}
\maketitle

\setcounter{figure}{0}
\setcounter{section}{0}

\makeatletter
\renewcommand{\fnum@figure}{Fig. S\thefigure}
\makeatother
\normalsize
\section{\label{sec:1} Microbubble synthesis and sample preparation}
A phospholipid shell was used to stabilize the microbubble which consisted of 1, 2-dipalmitoyl- sn -glycero -3- phosphocholine (DPPC) and 1, 2- distearoyl-sn- glycero-3- phosphoethanolamine -N- [methoxy (polyethylene glycol) – 2000] (DSPE - PEG2000) at a concentration of 2 mg/mL and a molar ratio 9:1 in phosphate buffer saline (PBS). High density perfluorobutane (PFB, C$_4$F$_{10}$), was used as the gas core of the microbubbles. The lipid solution and PFB gas interface was placed onto a probe tip sonication device at a constant, high output power for 45 seconds to form the microbubbles. Centrifugation was used to wash away residual lipid solution and to size isolate the microbubbles into different size distributions \cite{Feshitan}. The average of three measurements from a Multisizer Coulter Counter was used to characterize the size distribution and concentration of the microbubble distributions as shown in Fig. S2. Immediately before each photoacoustic experiment, 3 mL of ink and blood samples were mixed with the microbubbles at the desired concentration, keeping the 3 mL of volume constant. 

\section{\label{sec:2} Singular value decomposition}
Singular value decomposition (SVD) has been applied in numerous ultrasound techniques including clutter rejection in ultrafast Doppler \cite{Demene,Baranger} and localization-based super-resolution imaging \cite{Errico,ZhangAWSALM}. As we proposed a fluctuation-based imaging approach, SVD was required to separate unwanted fluctuation sources, such as laser fluctuations, that tend to dominate the fluctuation image. 

SVD processing followed the acquisition and delay-and-sum reconstruction of the 1,000 RF frames, where the in-phase and quadrature (IQ)  frames were written as a complex matrix $s(n_z,n_x,n_t)$, where  $n_z$ and $n_x$ correspond to the number of points in the $z$ and $x$ directions of the reconstructed IQ frames, respectively, and $n_t$ is the number of frames collected. We then reshaped the IQ frames into a complex 2D space-time matrix $\boldsymbol{S}$ with dimensions $(n_z\times n_x,n_t)$. According to the SVD, we can write $\boldsymbol{S}$ as:

\begin{equation}
    \boldsymbol{S=U\Sigma V^T}
\end{equation}

\noindent where $\boldsymbol{\Sigma}$ is a diagonal matrix of size $(n_t\times n_t)$ containing singular values ordered from largest to smallest, $\boldsymbol{U}$ and $\boldsymbol{V}$ are orthonormal matrices with respective dimensions $(n_z\times n_x,n_t)$ and $(n_t\times n_t)$. The columns of matrices $\boldsymbol{U}$ and $\boldsymbol{V}$ are the spatial and temporal singular vectors of $\boldsymbol{\Sigma}$, respectively. Once $\boldsymbol{S}$ is decomposed into these three matrices, we chose which singular components to keep in $\boldsymbol{\Sigma}$.

\newpage
%Fig1
\begin{figure}[b]
    \centering
    \includegraphics[width=\linewidth]{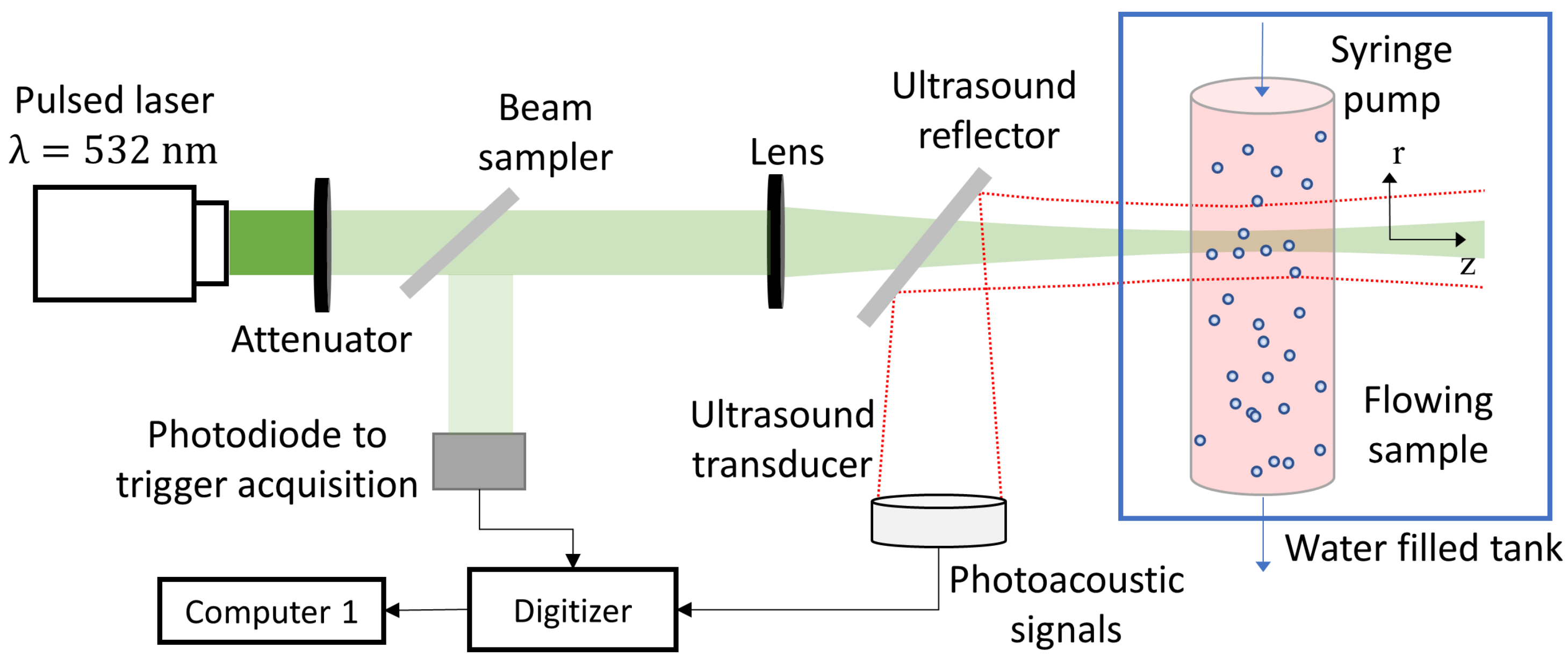}
    \caption{\label{fig:sExp1} Experimental setup for single element transducer measurements for $\sigma_N$ as a function of microbubble concentration and size distribution.}
\end{figure}

\begin{figure}[h!]
    \centering
    \includegraphics[width=\linewidth]{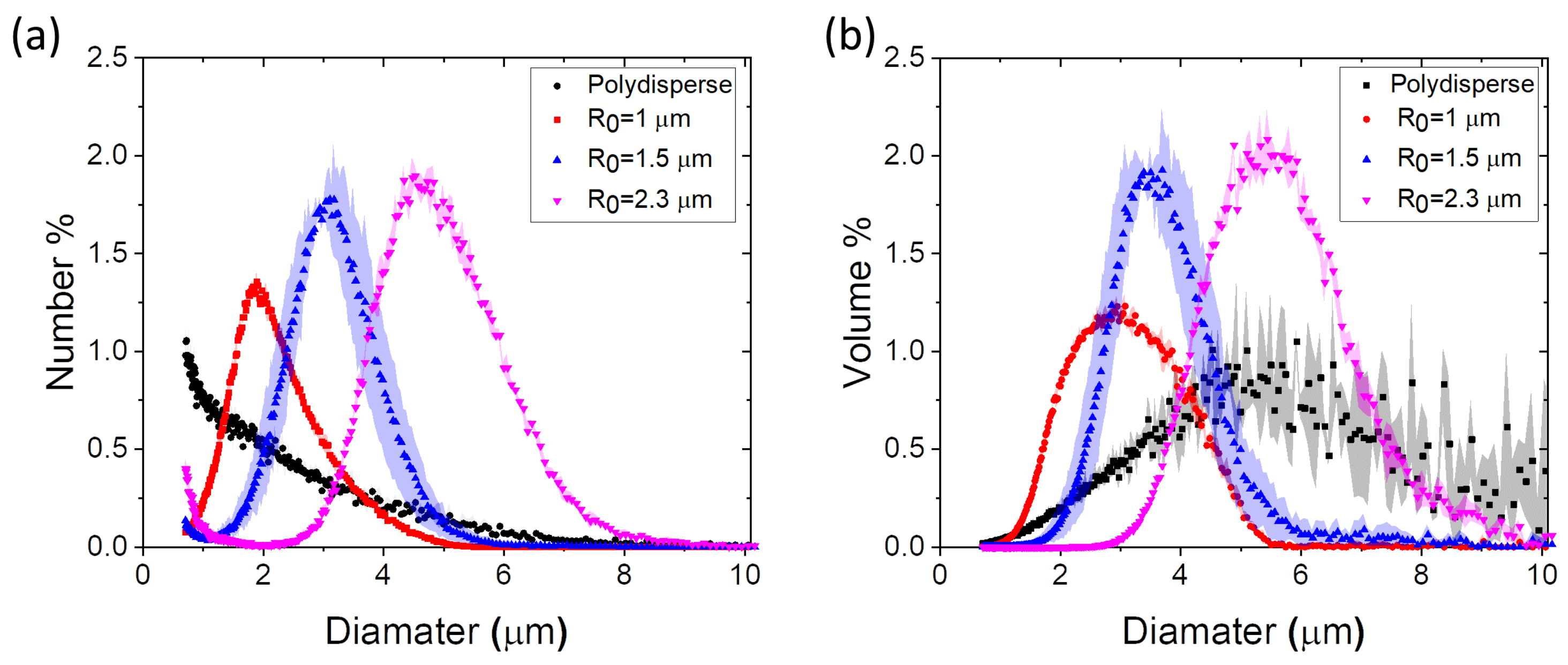}
    \caption{\label{fig:sExp2} Microbubble size distributions measured in a Multisizer Coulter Counter. Reporting mean and standard deviation of N = 3 measurements. (a) Number percent of three size isolated and polydisperse microbubble size distributions with different mean diameters. (b) Volume percent of the same microbubble size distributions.}
\end{figure}

\newpage
\begin{figure}[h!]
    \centering
    \includegraphics[scale=0.4]{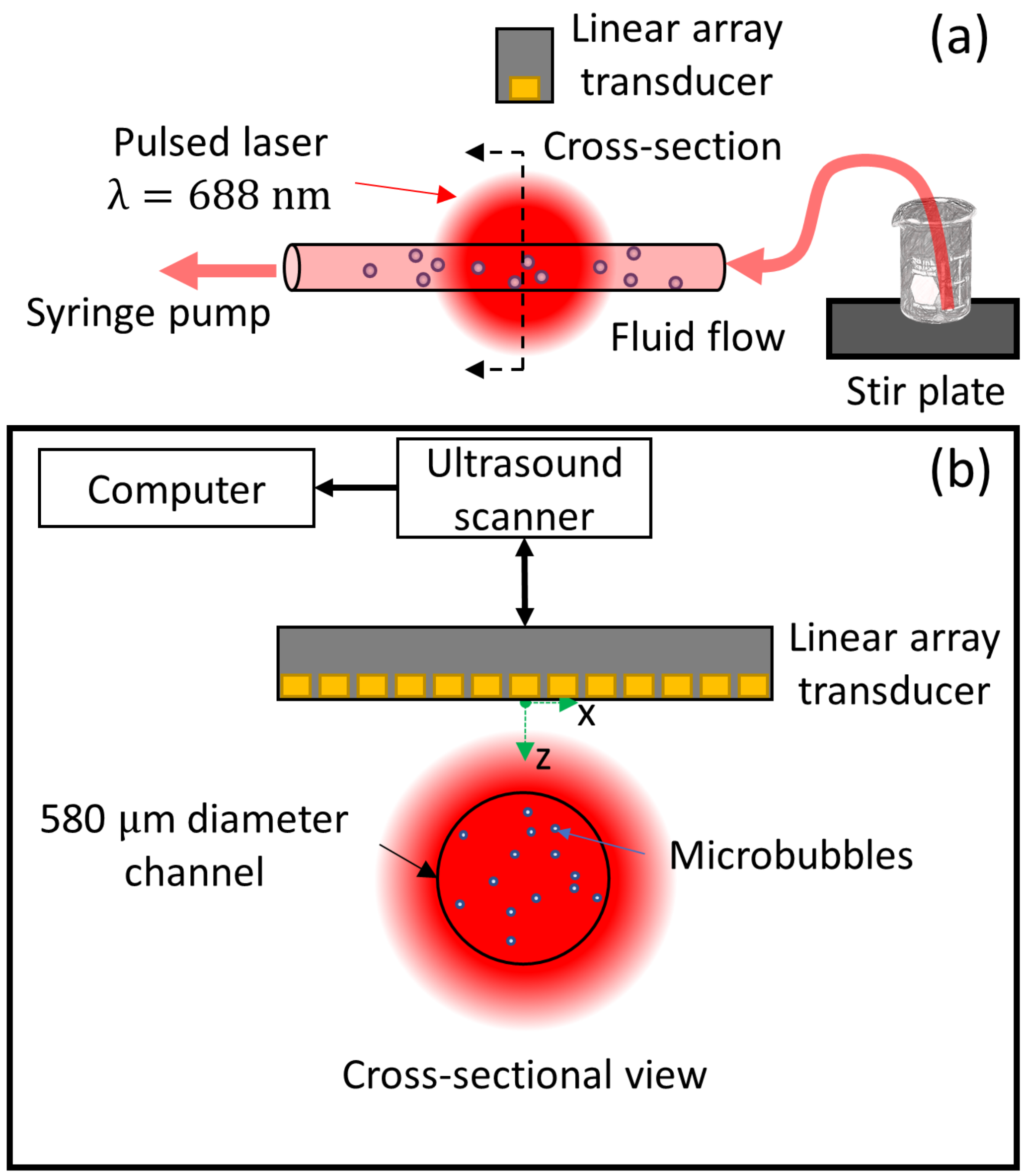}
    \caption{\label{fig:sExp3}  Experimental set-up for photoacoustic imaging. (a) A flow-phantom was illuminated by a pulsed-laser (7 ns pulses, $\lambda$=688 nm) and the photoacoustic signals were detected by a linear-array transducer (128 elements, 20 MHz center frequency, L22-14vx). A syringe pump was used to control the flow of the MB sample through the phantom. (b) Cross-sectional view of the phantom and acoustic detection.}
\end{figure}

\newpage
\begin{figure}[h!]
    \centering
    \includegraphics[width=\linewidth]{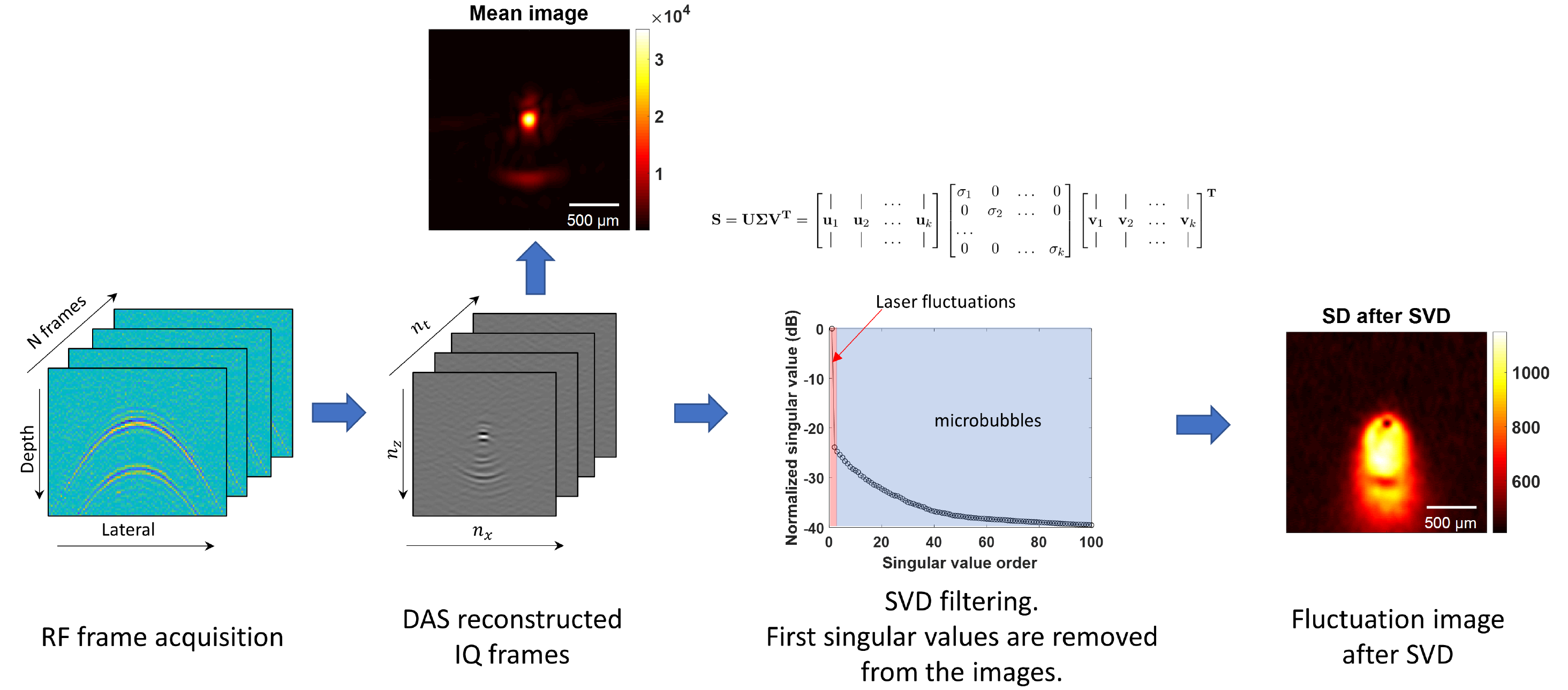}
    \caption{\label{fig:sExp4}  Photoacoustic fluctuation imaging flow diagram.}
\end{figure}

\end{document}